\RequirePackage{lineno}
\documentclass[pra,twoside,onecolumn,showpacs,aps,longbibliography]{revtex4-2}
\usepackage{amsmath}
\usepackage{amssymb}
\usepackage{amsfonts}
\usepackage{bm}
\usepackage{mathtools}
\usepackage{subfig}
\usepackage{graphicx}
\usepackage{datetime}
\usepackage{fancyhdr}
\usepackage{datetime}
\usepackage{array}
\usepackage{booktabs}
\usepackage{comment}
\usepackage{multirow}
\usepackage{geometry}
\usepackage{url}

\newcommand{\arcosh}{{\rm arcosh\,}}

\newcommand{\alphaC}{{\alpha_{\rm C}}}

\newcommand{\thetaC}{{\theta_{\rm C}}}

\fancypagestyle{plain}{
    \fancyhf{}
    \fancyfoot[C]{\today\ \currenttime}
    }

\DeclareMathAlphabet{\pazocal}{OMS}{zplm}{m}{n}

\begin{document}


\title{
Bi-Constructible pattern  of weak and flavour mixing: implications for electroweak coupling constants\\
}

\author{Jacek Ciborowski}
 \affiliation{Faculty  of Physics, University of Warsaw, L.~Pasteura 5, PL-02-093 Warsaw, Poland.}
  \email{cib@fuw.edu.pl}

\begin{abstract}
We present semi-empirical evidence suggesting that weak and flavour mixing, at the most fundamental level, can be described in terms of the Euclidean geometry of regular polygons constructible with compass and straightedge, specifically, the pentagon and the heptadecagon, associated with Fermat primes---a pattern referred to as  \emph{Bi-Constructible}.  Our approach accurately reproduces quark and lepton mixing angles and offers  indications that the Weinberg angle also fits naturally within this geometric framework.  Concise Weak--Quark--Lepton Complementarity  relations are derived. These findings  suggest a semi-empirical  unification pattern of weak and flavour mixing.  The Standard Model gauge couplings $g$ and $g^{\prime}$ admit elegant expressions involving the golden ratio, yielding a neat prediction for the fine-structure constant entirely in these terms.
\end{abstract}

\maketitle

\large

\section{Introduction}\label{sec:Introduction}

The phenomena of flavour mixing and CP violation play a central role in both the quark and  lepton  sectors,  describing   weak decays of hadrons as well as other flavour-changing hadronic processes and neutrino oscillations.
Despite significant advancements, our understanding of the origin and values of quark and lepton mixing angles, and   CP-violating phases, remains
incomplete. The flavour mixing processes are governed by  the the Cabibbo-Kobayashi-Maskawa (CKM) matrix for quarks~\cite{cite:CKM,cite:CKM2,cite:PDG2024} and the
Pontecorvo-Maki-Nakagawa-Sakata (PMNS) matrix for leptons~\cite{cite:PMNS}. These unitary  matrices encode  respective mixing angles and CP-violating phases and both arise from the same phenomenon  in the Standard Model that  weak interaction eigenstates differ from mass eigenstates.
In this paper we  denote and order the   three mixing angles in each sector  by:  $\theta_{ij}^{q,l},\;ij=12,23,13$, where the labels  '$q$' and '$l$'  refer to   quarks and leptons, respectively.

In  this analysis we use the experimental values   presented in Table~\ref{tab:MixingExp} and labelled as 'exp'.
As for quarks, these are the best-fit values from global fits to current  measurements of the CKM matrix elements~\cite{cite:PDG2024} (2024). For neutrinos,  normal
ordering values were taken from the best-fit results of  global fits  NuFit-6.0~\cite{cite:NuFit2020,cite:NuFit2024} (2024), with exclusion of $\theta_{23}^{l}$, as explained below.  NuFit-6.0 provided the best-fit value  $(48.5^{+0.7}_{-0.9})^{\circ}$,  obtained with exclusion of the Super Kamiokande (SK)
atmospheric data,  derived on the basis of  the  slightly lower  of the two narrow  local minima  of the  $\chi^2$ vs. $\sin^2\theta_{23}^{l}$
dependence. It  is characterised by  unprecedentedly  small $1\,\sigma$-uncertainties, whereas   already  the  $2\,\sigma$ limits point to  a very  broad  range  of values,  approximately  $(40\,\text{--}\,50)^{\circ}$. A significantly smaller  best-fit  value, with equally small standard deviations,  $(43.3^{+1.0}_{-0.8})^{\circ}$, was obtained for the dataset  in which  the SK atmospheric data remained included.
In this context,  concerns may be raised about the choice of a particular value  and the potential underestimation of its associated uncertainties. We  have thus  adopted the recent IceCube measurement~(2023)~\cite{cite:Abbasi1}, $\theta_{23}^{l({\rm exp})}=(45.6\pm 2.9)^{\circ}$.
It offers a more comprehensive representation of the ambiguities and variability inherent in this dataset,  even though the  associated uncertainty of this conservative estimate remains large and  somewhat weakens the foregoing analysis.
The measurement of IceCube remains  in an  agreement with results of  other recent  analyses:  $[48.1, 49.5]^{\circ}$~($1\,\sigma$ range)~\cite{cite:Arguelles1},  $\left(47.9^{+1.1}_{-4.0}\right)^{\circ}$~\cite{cite:Capozzi2},  and measurements:     $(49.0^{+2.6}_{-3.5})^{\circ}$~(NOvA 2023)~\cite{cite:Acero1} and  $\left(48.5^{+1.2}_{-1.8}\right)^{\circ}$~(T2K 2023)~\cite{cite:Abe1}.
In the case of asymmetric errors we have adopted the larger one as  the symmetric uncertainty~(Table~\ref{tab:MixingExp}) for the sake of simplicity of presentation.

\begin{table}[ht]
    \centering
    \begin{tabular}{lccc}
\toprule
  Mixing   angle  &  $\theta^{({\rm exp})}_{12}$   &      $\theta^{({\rm exp})}_{23}$  &        $\theta^{({\rm exp})}_{13}$     \\
\midrule
Quarks  & $13.004\pm0.040 $& $2.397^{+0.045}_{-0.040} $ &  $0.2138^{+0.0052}_{-0.0049}$ \\
\midrule
Leptons   &  $ 33.68^{+0.73}_{-0.70}$ &  $ 45.6 \pm 2.9  $  &   $  8.56\pm 0.11 $ \\
\midrule\midrule
$\theta_{\rm W}$ & & $(28.1931\pm 0.0182)^{\circ}$ &\\
$\sin^2 \theta_{\rm W}$ &&  $0.2232 \pm 0.0003$\\
$M_{\rm W}$ &    & $80.3692\pm 0.0133$~GeV &\\
$M_{\rm Z}$ &    & $91.1880\pm 0.0020$~GeV &\\
$M_{\rm W}/M_{\rm Z}$ &    & $0.88136 \pm 0.00015$ &\\
\bottomrule
\end{tabular}
 \caption{Best-fit values of the quark~\cite{cite:PDG2024}  and lepton~(NO)~\cite{cite:NuFit2020,cite:NuFit2024}  mixing angles (degrees),  obtained from  global fits, with exception of the atmospheric angle (see text). The  value of the 'on shell'  Weinberg angle, $\theta_{\rm W}$, is evaluated  from the  SM relation $\cos \theta_{\rm W}=M_{\rm W}/M_{\rm Z}$~\cite{cite:PDG2024}. }
    \label{tab:MixingExp}
\end{table}

The dominant  approach to deriving flavour  mixing angles relies on  flavour symmetries and Grand Unified Theories (GUTs).
The predictions of the former are typically obtained  by imposing constraints on the Yukawa couplings.
The  most commonly considered   discrete symmetries include  $S_4$, $A_4$, $A_5$, $D_{10}$,
and $T^{\prime}$~\cite{cite:Ishimori1,cite:King1,cite:Everett1,cite:Babu1,cite:Barman1,cite:Ghosh1,cite:Altarelli3,cite:Morisi1,
cite:Kim1,cite:Bazzocchi1,cite:Ma3,cite:Feruglio1} (see~\cite{cite:Petcov2,cite:Meloni1} for reviews).
Among these, $A_5$ and $D_{10}$ have received particular attention for their connection with mixing patterns involving the golden ratio, $\varphi$,
being the positive solution of the equation~(\ref{eq:GoldenEquation}a)
\begin{equation}\label{eq:GoldenEquation}
 \varphi^2 - \varphi - 1 = 0 \;\;\; ({\rm a}) \qquad  \qquad   b^{({\rm G})}=1/\varphi^2 \approx 0.3820 \;\;\; ({\rm b}),
\end{equation}
equal to  $\varphi = (1 + \sqrt{5})/2  \approx 1.6180$.  Dividing unity, or a unit-length line segment,  according to the golden proportion yields two complementary fractions of which the smaller is given by~(\ref{eq:GoldenEquation}b)
and the larger by  $1- b^{({\rm G})}=1/ \varphi \approx 0.6180$.

The icosahedral group $A_5$ is notable for predicting lepton mixing angles, albeit only in rough agreement with experimental data, e.g.,  $\tan \theta_{12}^{l} = 1/\varphi$, which corresponds to $\theta_{12}^{l} \approx 31.7^{\circ}$~\cite{cite:Everett1,cite:Kajiyama1}.
The dihedral group $D_{10}$, associated with the rotational and reflection symmetries of a regular decagon, yields specific round numbers, e.g.,  $\cos \theta_{12}^{l} = \varphi/2$~\cite{cite:Rodejohann1,cite:Adulpravitchai1}, giving $\theta_{12}^{l} = 36^{\circ}$, equal to the exterior angle of a regular decagon. While both values lie near the observed solar mixing angle, they fall outside the current best-fit range, highlighting the limitations of such symmetry-based predictions.

While models based on flavour symmetries remain the most widely studied, several alternative or complementary frameworks have also been proposed.
In the minimal flavour violation  framework it is  postulated that all sources of flavour violation---including those potentially arising from new physics---are governed by the Standard Model Yukawa couplings~\cite{cite:DAmbrosio1,cite:Kagan1,cite:Smith1}.
In  geometric and extra-dimensional approaches, hierarchical flavour structures arise from the localisation of fermions in extra spatial dimensions, with effective Yukawa couplings generated without invoking explicit family symmetries~\cite{cite:ArkaniHamed1}.
Some flavour models attribute mixing to vacuum expectation values (VEV) of flavour-breaking scalars (flavons) that  break the original flavour group~\cite{cite:Varzielas1}.
Predictions closer to the present energy scale have been obtained in hierarchical VEV models,  in which the neutrino mixing angles are expressed as functions of the Cabibbo angle and  also by  the strange and charm quark masses, yielding predictions that approximately  reproduce the lepton data:
$\theta_{12}^{l} \approx 33.37^{\circ}$, $\theta_{23}^{l} \approx 50.81^{\circ}$, and $\theta_{13}^{l} \approx (8.12$--$8.68)^{\circ}$~\cite{cite:Abbas1,cite:Abbas2}.
Generative artificial intelligence and numerical scans of parameter space have also been employed in  data-driven approaches to identify phenomenologically viable flavour parameters, even in the absence of symmetry constraints. Such  methods offer a complementary tool for exploring flavour models~\cite{cite:Kobayashi2}.
Flavour mixing angles have also been proposed based on ad hoc simplicity criteria. Examples include identifying the Cabibbo angle, $\theta_{12}^{q}$, and the solar angle, $\theta_{12}^{l}$, with rational fractions of the straight angle, such as $15^{\circ}$ and $36^{\circ}$, respectively; similarly, the reactor angle has been suggested to take the value $9^{\circ}$~\cite{cite:Rodejohann1,cite:Rodejohann2}.

In Grand Unified Theories (GUTs), typically based on gauge groups such as $SU(5)$ or $SO(10)$, quarks and leptons are embedded into unified multiplets. These theories aim not only to unify the fundamental interactions under a single gauge group, but also to provide a structural explanation for observed regularities in the flavour sector, offering a natural framework for correlating mass and mixing patterns~\cite{cite:Raidal2,cite:Antusch1} (see~\cite{cite:King6} for a review). The breaking of the GUT symmetry down to the Standard Model gives rise to specific relationships between quark and lepton mixing angles within a unified theoretical setting.  To obtain realistic fermion mass hierarchies and mixing structures, GUTs are often supplemented by flavour symmetries, which constrain the form of the Yukawa couplings and mass matrices. These structural constraints can lead to testable relations among mixing angles, most notably in the form of Quark--Lepton Complementarity (QLC), where the absolute values of quark and lepton mixing angles obey approximate sum rules, such as $\theta_{12}^{l} + \theta_{12}^{q} \approx \theta_{23}^{l} + \theta_{23}^{q} \approx 45^{\circ}$~\cite{cite:Raidal2,cite:Schmidt1,cite:Zhang1,cite:Zhukovsky1}.
While early data on lepton mixing were suggestive of such underlying correspondences, they are not fully supported by current, higher precision measurements.

The above outline underscores that predictions derived from abstract symmetry group considerations remain partial---they neither exhibit robust agreement with present experimental results nor offer an exhaustive description, leaving substantial room for refinement and alternative approaches.

Among the key parameters of the electroweak sector of the Standard Model is the Weinberg angle, $\theta_{\rm W}$, also referred to as the weak mixing angle. It quantifies the mixing between the $\mathrm{SU}(2)_L$ and $\mathrm{U}(1)_Y$ gauge fields, ultimately giving rise to the physical $Z$ and (massless) photon fields observed in Nature. The Weinberg angle plays a central role in determining the relative strengths of the weak and electromagnetic interactions, and its precise value influences numerous electroweak processes. It remains an input parameter of the Standard Model, with its numerical value determined empirically rather than derived from first principles. In our analysis, we adopt the on-shell definition of the Weinberg angle, $\cos \theta_{\rm W} = M_{\rm W}/M_{\rm Z}$, where $M_{\rm W}$ and $M_{\rm Z}$ are the masses of the electroweak vector bosons~\cite{cite:PDG2024} (Table~\ref{tab:MixingExp}).

The overarching aim of the present study is to search for regularities or patterns hidden in the 'small' data of measured mixing angles, spanning both the electroweak and flavour sectors. This pursuit  complements symmetry-based theoretical frameworks, viewed from the perspective of the current energy scale of the Universe. We proceed under the assumption that the relevant structure can be captured in terms of simple constructs.

We show that the  flavour mixing angles can be accurately predicted from a number-theoretical and geometrical framework based on two constructible regular polygons—those that can be drawn using compass and straightedge—associated with Fermat primes: the regular pentagon (linked to the golden ratio) and the regular heptadecagon (associated with it, up to subtleties). We refer to this pattern as \emph{Bi-Constructible}. Our formalism  does not involve neither quark nor lepton masses. The Weinberg angle fits naturally into the same scheme, allowing the Standard Model gauge couplings and the fine-structure constant to be expressed in terms of the golden ratio. The resulting predictions match experimental values with very good precision and yield concise Weak--Quark--Lepton Complementarity (WQLC) relations.
CP-violating phases are not addressed, as the limited precision of measurements in the lepton sector prevents a meaningful empirical analysis.

\section{Preliminaries}\label{sec:Preliminaries}

The constructibility of regular polygons was  formalised in a theorem by Wantzel~\cite{cite:Wantzel} (1837), following earlier contributions by Euclid~(triangle, pentagon)~\cite{cite:Euclides} and Gauss~(heptadecagon)~\cite{cite:Gauss}. According to Wantzel's result, a regular polygon with $n$ sides, denoted $P_{n}$, can be constructed using  compass and straightedge if and only if
\begin{equation}\label{eq:Wantzel}
n = 2^k \cdot p_{1} \cdot p_{2} \cdots p_{m},
\end{equation}
where $k$ is a non-negative integer and the $p_i$ are distinct Fermat primes. A Fermat number is  of the form $F_{m} = 2^{2^m} + 1$, where $m$ is a non-negative integer. To date, only five Fermat numbers are known to be prime: $F_0 = 3$, $F_1 = 5$, $F_2 = 17$, $F_3 = 257$, and $F_4 = 65537$.
The second and third Fermat primes correspond to the constructible regular pentagon and heptadecagon---precisely the geometric figures that arise from the present analysis of flavour mixing angles.

The cornerstone of our formalism is the idea of considering  pairs of mixing angles in each sector. This forms the basis of the \emph{polygonal scheme}, which reduces the investigation to smaller, more tractable subsystems. The pairs  are ordered as follows:  $12-23$, $23-13$ and $12-13$ and indexed by $i=1,2,3$, respectively. For each pair, a right-angled \emph{mixing triangle} is constructed, in which the two mixing angles are identified with its  catheti. These triangles are rooted in geometric relationships with regular $n$-sided polygons, yielding a two-dimensional representation. Each mixing angle in a given pair is classically parametrised using polar coordinates: a common acute angle of the triangle, $\alpha_{i}$, referred to as the \emph{mixing seed (angle)}, and the hypotenuse, $R_{i}$, referred to as the \emph{mixing strength}. The  parametrisations   read explicitly
\begin{align}
\theta_{12} &= R_{1} \cos\alpha_1  &&\hspace{-1em} ({\rm a}) &
\theta_{23} &= R_{1} \sin\alpha_1  &&\hspace{-1em} ({\rm b})  \label{eq:ParamPolar1} \\[0.5em]
\theta_{23} &= R_{2} \cos\alpha_2  &&\hspace{-1em} ({\rm a}) &
\theta_{13} &= R_{2} \sin\alpha_2  &&\hspace{-1em} ({\rm b})  \label{eq:ParamPolar2} \\[0.5em]
\theta_{12} &= R_{3} \cos\alpha_3   &&\hspace{-1em} ({\rm a}) &
\theta_{13} &= R_{3} \sin\alpha_3  &&\hspace{-1em} ({\rm b}), \label{eq:ParamPolar3}
\end{align}
where the labels '$q$' and '$l$' have been omitted for notational clarity.
Since the relation $\tan\alpha_{1} \tan\alpha_{2} = \tan\alpha_{3}$ can be straightforwardly derived, it is sufficient to analyse only the first two pairs of mixing angles in a given sector. This approach allows the study of flavour mixing to be recast in terms of the parameters $R_i$ and  $\alpha_i$, $i = 1, 2$,  rather than the mixing angles themselves.

The quark mixing angle $\theta_{12}^{q}$ is commonly referred to as the Cabibbo angle and is denoted below by $\theta_{\rm C}$. The quark flavour mixing seed in the first quark pair, $\alpha_{1}^{q}$, will be termed the \emph{Cabibbo mixing seed} and denoted by $\alpha_{\rm C}$, while the corresponding mixing strength, $R_{1}^{q}$---$R_{\rm C}$. The Cabibbo mixing seed, once assigned a specific numerical value, together with the golden or para-golden ratio (see below), will play a central role in the present analysis. Experimental values of $R_{i}^{({\rm exp})}$ and $\alpha_{i}^{({\rm exp})}$, $i = 1, 2$,  are presented in Table~\ref{tab:Quantities}.

\begin{figure}[!htbp]
        \includegraphics[width= 8.0cm]{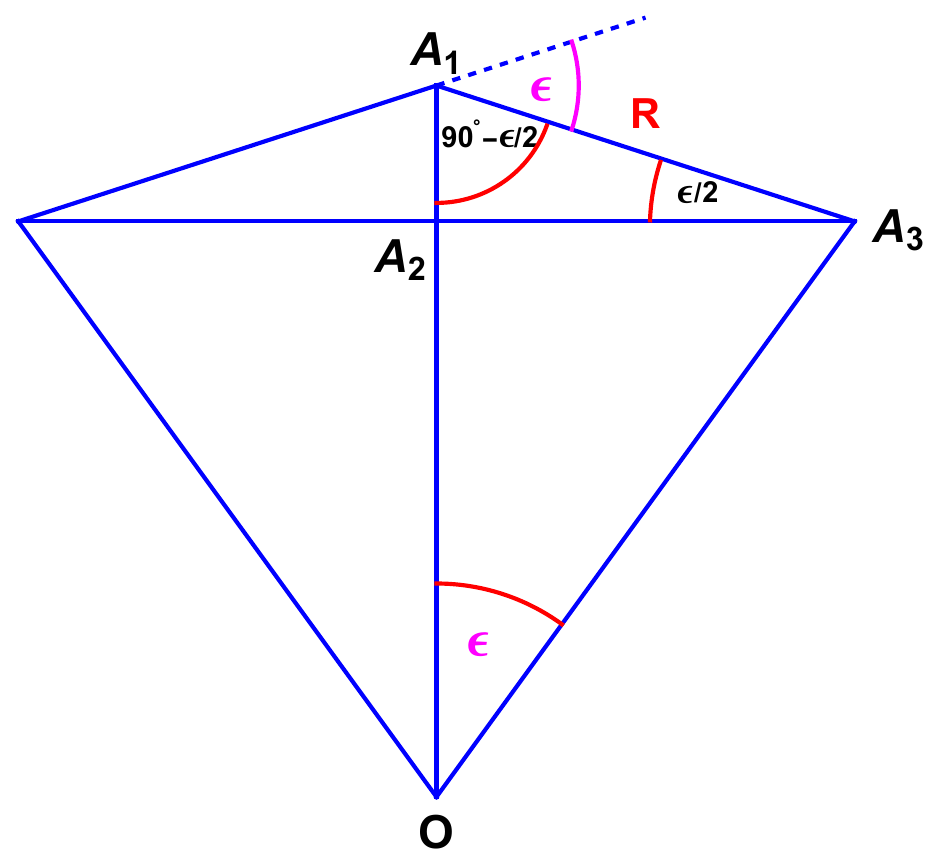} 
    \caption{\label{fig:PolygonGeneral} Geometry of a regular polygon visualised on two adjacent sectors. The exterior angle $\epsilon$ equals $360^{\circ}/n$. The mixing triangle is identified as $\triangle A_1A_2A_3$ with the hypothenuse coinciding with  the polygon side of length corresponding to the mixing strength, $R$.}
\end{figure}

Two adjacent sectors of a regular polygon are shown in Fig.~\ref{fig:PolygonGeneral}. The mixing triangle is identified with  $\triangle A_1A_2A_3$, and the angle $\angle A_2A_3A_1$ equals one-half of the exterior angle of the polygon, $\epsilon$,   related to the rank of the polygon, $n$, by
\begin{equation}\label{eq:ExteriorAngle}
\epsilon = \frac{360^{\circ}}{n}.
\end{equation}
Each mixing triangle  is uniquely defined by the number of sides and the side length of the corresponding regular polygon. Since the hypotenuse of a given mixing triangle coincides with a side of the polygon, the mixing strength $R_i$ is naturally associated with the polygon's side length.

We also introduce the \emph{normalised mixing angles} within each pair. These are respectively labelled 'n1' and  'n2'.  A normalised mixing angle in a given pair  is defined as a mixing angle divided by the mixing strength of the corresponding pair and thus  it equals the sine or cosine of the associated mixing seed angle. Equivalently, a given pair of normalised mixing angles is represented by a mixing triangle with unit hypotenuse.  These angles explicitly read
\begin{align}
\theta_{12({\rm n1})} &=  \frac{\theta_{12} }{ R_{1}} = \cos\alpha_1 &&\hspace{-1em} ({\rm a})
&   \theta_{23({\rm n1})} = \frac{ \theta_{23} }{ R_{1}} =  \sin\alpha_1  &&\hspace{-1em} ({\rm b})\label{eq:DefNormal1} \\
\theta_{23({\rm n2})} &=  \frac{\theta_{23} }{ R_{2}} = \cos\alpha_2   &&\hspace{-1em} ({\rm a})
&   \theta_{13({\rm n2})} = \frac{\theta_{13} }{ R_{2}} =  \sin\alpha_2 &&\hspace{-1em} ({\rm b}). \label{eq:DefNormal2}
\end{align}
In particular, the angle $\theta^{q}_{12({\rm n1})}$ will be denoted by $\theta_{\rm C({\rm n})}$.

For the solar and atmospheric angles, an alternative scheme, referred to as the \emph{fractional scheme}, is also available, rendering the description dual for this specific pair. In this scheme, each of the two angles arises from the division of a line segment into two  unequal parts according
to a rule defined by the golden ratio~(Sec.~\ref{sec:Leptons}).
A similar concept of division---though governed by a different rule---also appears in the context of weak mixing.

Throughout this work, we use the term \emph{$\varepsilon$-equivalent}, and similarly $\varepsilon$-prefixed variants, to denote agreement of mathematical expressions at the level of a per mille or better. The formulation of the present framework is intrinsically linked to, and contingent upon, these quasi-equalities.

\section{Lepton mixing}\label{sec:Leptons}

We begin the analysis by identifying the values of the solar and atmospheric mixing angles, $\theta_{12}^{l}$ and $\theta_{23}^{l}$, within the polygonal scheme. The experimental values of the relevant quantities for the lepton pair $12$--$23$ are listed in Table~\ref{tab:Quantities}. In this framework, the predicted  mixing angles, hereafter labelled 'P5', are represented by the catheti $A_1A_2$ and $A_2A_3$ of the mixing triangle~(Fig.~\ref{fig:PolygonGeneral}).
The measured seed angle, $(53.6\pm 1.8)^{\circ}$,  can be   identified with the angle $\angle A_2A_1A_3 = 54^{\circ}$ corresponding to the regular pentagon $P_5$, which is expressed by the respective exterior angle, $\epsilon=72^{\circ}$
\begin{equation}\label{eq:DefP5Seed}
   \alpha_{\rm{P5}} := \alpha_1^{l({\rm P5})} = 90^{\circ} - \epsilon /2= 54^{\circ}.
\end{equation}
The experimental value of $R_1^{l({\rm exp})}$ is consistent with  unit mixing strength, and thereby supports  assigning  $R_1^{l({\rm P5})} = 1$ in this scheme---a point that will be further substantiated below.
The regular pentagon is a constructible polygon, associated with the second Fermat prime, $F_1 = 5$.

\begin{table}[ht]
    \centering
    \begin{tabular}{lccc}
\toprule[1pt]
&& Pair $\theta_{12}$—$\theta_{23}$& \\
\midrule[1pt]
       & $R_{1}^{({\rm exp})}$ & $\alpha_{1}^{({\rm exp})}$  &  $n_{1}^{({\rm exp})}$    \\
       & (rad) &  (deg) &    \\
\midrule
Quarks  & $0.23079\pm 0.00070$ & $10.444\pm 0.194$& $17.2\pm 0.3$    \\
\midrule
Leptons & $0.99\pm 0.04 $ & $53.6\pm 1.8 $ &   $4.9 \pm 0.3$    \\
\toprule[1pt]
&& Pair $\theta_{23}$—$\theta_{13}$& \\
\midrule[1pt]
       & $R_{2}^{({\rm exp})}$ & $\alpha_{2}^{({\rm exp})}$  &  $n_{2}^{({\rm exp})}$     \\
       & (rad) &  (deg) &     \\
\midrule
Quarks  & $0.0420\pm 0.0008$ & $5.097\pm 0.156$& $35.3\pm 1.1$   \\
\midrule
Leptons & $0.810\pm 0.050 $ & $10.6\pm 0.7 $ &   $16.9 \pm 1.1$    \\
\bottomrule[1pt]
\end{tabular}
\caption{  Experimental  values of quantities relevant to  this study,   obtained   using data  from Table~\ref{tab:MixingExp} for the 1st and 2nd   pair of mixing angles, as defined by~\eqref{eq:ParamPolar1}--\eqref{eq:ParamPolar2}.}
 \label{tab:Quantities}
\end{table}

According to the above identification and adopting  a one-unit mixing strength, the first two lepton flavour mixing angles  are given by
\begin{equation}\label{eq:DefP5}
\theta_{12}^{l({\rm P5})} = \cos \alpha_{\mathrm{P5}} = \sqrt{1 - \left(\frac{\varphi}{2}\right)^2} \approx 33.6776^{\circ} \;\; ({\rm a}) \quad
\theta_{23}^{l({\rm P5})} = \sin \alpha_{\mathrm{P5}} = \frac{\varphi}{2} \approx 46.3533^{\circ} \;\; ({\rm b}).
\end{equation}
The predicted value of the solar angle agrees remarkably well with the measured value, within its moderate uncertainty; idem the atmospheric angle, albeit within the  uncertainty adopted conservatively~(Table~\ref{tab:MixingExp}).

We now specify the  alternative, fractional scheme for defining the first two lepton mixing angles, subsequently labelled 'G'. This scheme arises naturally from the following, nonobvious  quasi-equalities involving the golden ratio
\begin{equation}\label{eq:PuzzlingQuasiEpsilon}
\frac{1}{2} \arccos\left(\frac{1}{\varphi^2}\right) =_{\varepsilon} \sqrt{1 - \left(\frac{\varphi}{2}\right)^2} \;\; ({\rm a}) \qquad
\frac{1}{2} \arcosh\left(\varphi^2\right) =_{\varepsilon} \frac{\varphi}{2} \;\; ({\rm b}),
\end{equation}
which hold at a  per mille level. Consequently, we define  the mixing  angles in the 'G' scheme  by the left-hand sides of Eqs~\eqref{eq:PuzzlingQuasiEpsilon} and  thus get
\begin{equation} \label{eq:DefGolden}
\cos\left(2\theta_{12}^{l({\rm G})}\right) = \frac{1}{\varphi^2} = b^{({\rm G})} \approx 0.3820 \;\; ({\rm a}) \quad
\cosh\left(2\theta_{23}^{l({\rm G})}\right) = \varphi^2 = B^{({\rm G})} \approx 2.6180 \;\; ({\rm b}).
\end{equation}
The numerical values of the first two lepton mixing angles following from Eq.~\eqref{eq:DefGolden}
\begin{equation}\label{eq:DefG}
\theta_{12}^{l({\rm G})} \approx 33.7722^{\circ} \;\;({\rm a}) \qquad
\theta_{23}^{l({\rm G})} \approx 46.3214^{\circ} \;\;({\rm b})
\end{equation}
are nearly equal to their respective 'P5' counterparts~\eqref{eq:DefP5}, and are consistent with experimental data to a similar degree.

These results suggest introducing  two parameters, $b^{({\rm G})}$ and $B^{({\rm G})}$, where the former will be referred to as the \emph{small fraction parameter}  and, upon generalisation, will play an important role in subsequent considerations. From Eq.~\eqref{eq:DefGolden}, the following relations hold
\begin{equation}\label{eq:GoldenRelations3}
b^{({\rm G})} + B^{({\rm G})} = 3 \;\; ({\rm a}) \qquad
b^{({\rm G})} \, B^{({\rm G})} = 1 \;\;({\rm b}).
\end{equation}

We can evaluate the left-hand sides of Eq.~\eqref{eq:DefGolden} using the 'P5' values of the mixing angles, yielding slightly different results from their 'G' counterparts
\begin{equation}\label{eq:GoldenProportionP5}
\cos\left(2\theta_{12}^{l({\rm P5})}\right) = b^{({\rm P5})} \approx 0.3850 \;\;({\rm a}) \qquad
\cosh\left(2\theta_{23}^{l({\rm P5})}\right) = B^{({\rm P5})} \approx 2.6207 \;\;({\rm b}).
\end{equation}

Both small fraction parameters $b^{({\rm G})}$ and $b^{({\rm P5})}$ are expressed solely in terms of the golden ratio. We refer to the slightly modified division of unity into the fractions $b^{({\rm P5})}$ and $1 - b^{({\rm P5})} \approx 0.6150$ as the \emph{para-golden} division, to indicate its close but not exact correspondence to the golden division defined by $b^{({\rm G})}$ and $1 - b^{({\rm G})}$.

In the 'G' scheme, the solar mixing angle arises indirectly from assigning the trigonometric function in Eq.~\eqref{eq:DefGolden}a to the smaller fraction obtained by dividing a line segment of unit length in a golden  proportion. Both pairs of formulae, Eqs~\eqref{eq:DefP5} and~\eqref{eq:DefGolden}, confirm $\varepsilon$-equivalently  the role of the golden ratio in defining the first two lepton mixing angles.

While the polygonal mixing angles strictly satisfy the Pythagorean identity, those in the fractional scheme obey it with $\varepsilon$-level accuracy
\begin{equation}\label{eq:PuzzlingPythagorean}
\left[ \left( \theta_{12}^{l({\rm G})} \right)^2 + \left( \theta_{23}^{l({\rm G})} \right)^2 \right]^{1/2}
= \left[ \left( \tfrac{1}{2} \arccos\left( \tfrac{1}{\varphi^2} \right) \right)^2
+ \left( \tfrac{1}{2} \arcosh\left( \varphi^2 \right) \right)^2 \right]^{1/2}
=_{\varepsilon} 1,
\end{equation}
where the departure from unity amounts to  half a per mille.

The experimental values of the parameters $b^{({\rm exp})}$ and $B^{({\rm exp})}$ are
\begin{align} \label{eq:cos2thetaExp}
\cos 2\theta_{12}^{l({\rm exp})} = b^{l({\rm exp})} = 0.385 \pm 0.024 \;\;({\rm a}) \quad
\cosh 2\theta_{23}^{l({\rm exp})} = B^{l({\rm exp})} = 2.56 \pm 0.24 \;\;({\rm b}).
\end{align}

Substituting these values into the relations of Eq.~\eqref{eq:GoldenRelations3} yields
\begin{equation}\label{eq:GoldenRelations3exp}
b^{l({\rm exp})} + B^{l({\rm exp})} = 2.94 \pm 0.24 \;\;({\rm a}) \qquad
b^{l({\rm exp})} \, B^{l({\rm exp})} = 0.98 \pm 0.11 \;\;({\rm b}).
\end{equation}

Both the 'P5' and 'G' predictions are consistent with these values. Parenthetically, Eqs~\eqref{eq:DefP5} and~\eqref{eq:DefGolden} express the mixing angles in terms of the golden ratio in a form distinct from any  expressions previously proposed in literature (Sec.~\ref{sec:Introduction}).

It is natural to expect that the polygonal and fractional descriptions  form  a dual picture of mixing and thus  should yield $\varepsilon$-equivalent results for the values of the relevant mixing angles. This consistency requires that the side length of the pentagon be $\varepsilon$-equal to unity, as previously adopted. The first two lepton mixing angles in the 'P5' scheme are constructible, given  $R_1^{l({\rm P5})} = 1$, in contrast to those in the 'G' scheme, which are not. In both schemes, the angle $\theta_{23}^{l}$ lies in the second  octant.

The reactor angle $\theta_{13}^{l}$ can be predicted by identifying the regular polygon associated with the flavour mixing seed, $\alpha_{2}^{l}$, in  the second lepton pair,  $23$--$13$. The experimental values of quantities related to this pair  are summarised in Table~\ref{tab:Quantities}.

The experimentally determined  lepton mixing seed angle, $\alpha_{2}^{l({\rm exp})} = 10.6 \pm 0.7^{\circ}$,  can be  identified with the angle $\angle A_2A_3A_1$~(Fig.~\ref{fig:PolygonGeneral}). Owing to the sizeable uncertainty, three possible polygonal origins lie within the approximate $1\,\sigma$ range: $P_{16}$, $P_{17}$, and $P_{18}$~\eqref{eq:ExteriorAngle}, which correspond to $11.25^{\circ}$, $10.5882^{\circ}$ and $10^{\circ}$, respectively.
However, the choice can be guided by additional assumptions that allow for a more precise determination of the atmospheric angle within the present framework,  based on the measured solar angle. One may invoke the relation implied by the pentagon, $\theta_{23}^{l} = \theta_{12}^{l} \tan \alpha_{\rm P5}$, or alternatively, use the Pythagorean identity under the assumption of unit mixing strength as the constraint for this angle pair~\eqref{eq:ParamPolar1}.  This yields a significantly reduced uncertainties, respectively
\begin{equation}\label{eq:Theta23Prime}
\theta_{23}^{l({\rm exp}^{\prime})} = (46.36 \pm 1.00)^{\circ} \;\;({\rm a}) \quad
\theta_{23}^{l({\rm exp}^{\prime\prime})} = (46.35 \pm 0.53)^{\circ}  \;\;({\rm b}).
\end{equation}
Using these refined values, we obtain  $ \alpha_{2}^{l({\rm exp}^{\prime})} = (10.46 \pm 0.26)^{\circ}$~(a) and   $ \alpha_{2}^{l({\rm exp}^{\prime\prime})} = (10.46 \pm 0.18^{\circ})$~(b), favouring the heptadecagon which we  adopt in further considerations. Nevertheless, this case highlights the importance of obtaining a reliable and more precise, independent  measurement of the atmospheric angle.

We thus define the angle,  referring to it as the \emph{Cabibbo mixing seed}
\begin{equation}\label{eq:DefCabibboSeed}
\alpha_{\rm C} := \alpha_{2}^{l} = \frac{\epsilon}{2}=\frac{180^{\circ}}{17} \approx 10.5882^{\circ},
\end{equation}
which will be soon shown  to align with the seed in the first quark mixing pair, justifying the nomenclature. Since $\theta_{13}^{l} = \theta_{23}^{l} \tan \alpha_{\rm C}$, the 'P5' reactor angle is predicted by
\begin{equation}\label{eq:Theta13C}
\theta_{13}^{l({\rm P5})} = \sin \alpha_{\rm P5}  \tan \alpha_{\rm C},
\end{equation}
yielding $\theta_{13}^{l({\rm P5})} \approx 8.66^{\circ}$, within $1\,\sigma$ of the experimental value.   Consequently, the mixing strength in the 2nd lepton pair, is given by $R_{2}^{l({\rm P5})} = \sin \alpha_{\rm P5} / \cos \alpha_{\rm C}$ and evaluates to  $\approx 0.823$, which  agrees with the measured value.
The angle given by~\eqref{eq:Theta13C} is constructible.

Interestingly, the following nontrivial trigonometric $\varepsilon$-equality, accurate to better than one per mille, links the pentagon and the heptadecagon via the respective mixing seeds, $\alpha_{\rm P5}$ and $\alpha_{\rm C}$
\begin{equation}\label{eq:Quasi3}
\cos \alpha_{\rm P5} =_{\varepsilon} \pi \tan \alpha_{\rm C}.
\end{equation}
Owing to Eq.~\eqref{eq:Quasi3}, all three 'P5' lepton mixing angles can be neatly expressed with $\varepsilon$-accuracy in terms of fractions  of the straight angle
\begin{equation}\label{eq:AnglesLeptonsPi}
\theta_{12}^{l({\rm P5})} =_{\varepsilon} \pi \tan \alpha_{\rm C}
\;\;({\rm a}) \quad
\theta_{23}^{l({\rm P5})} =_{\varepsilon} \pi \tan \alpha_{\rm P5}  \tan \alpha_{\rm C}
\;\;({\rm b}) \quad
\theta_{13}^{l({\rm P5})} =_{\varepsilon} \pi \tan \alpha_{\rm P5}  \tan^2 \alpha_{\rm C}
\;\; ({\rm c}).
\end{equation}

Suggested  by Eq.~\eqref{eq:AnglesLeptonsPi}, it is natural to introduce  a consistency parameter, referred to as  the \emph{lepton mixing ratio}
\begin{equation}\label{eq:Rho}
\rho_{\rm m}^{l} = \frac{\theta_{12}^{l} \, \theta_{23}^{l}}{\pi \, \theta_{13}^{l}}.
\end{equation}
This ratio evaluates to unity with $\varepsilon$-accuracy,  $\rho_{\rm m}^{l({\rm P5})} =_{\varepsilon} 1$, when  the 'P5' values of the lepton mixing angles are substituted. The corresponding experimental value is  $\rho_{\rm m}^{l({\rm exp})} = 1.00 \pm 0.07$,
with the uncertainty dominated by that of the atmospheric angle. Eliminating the atmospheric angle by means of~\eqref{eq:ParamPolar1},
tighter bounds can be  obtained,   $ \rho_{\rm m}^{l({\rm exp}^{\prime})} = 1.01 \pm 0.05$ and  $\rho_{\rm m}^{l({\rm exp}^{\prime\prime})} = 1.01 \pm 0.02$, respectively, cf.~\eqref{eq:Theta23Prime}.

\section{Quark mixing}

Lacking an evident mathematical relation that would constrain the quark mixing strength within the first quark pair, $R_{\rm C}$ (which in the lepton sector arises from the fractional description),  we turn to empirical considerations in order to identify a possible quantitative connection between the two sectors.
We thus observe that the experimental value of the ratio
\begin{equation}\label{eq:Ratio12Exp}
b^{ql({\rm exp})} = \frac{\theta_{\rm C}^{({\rm exp})}}{\theta_{12}^{l({\rm exp})}} = 0.386 \pm 0.008
\end{equation}
is consistent with the experimentally determined value of the small fraction parameter $b^{l({\rm exp})}$ from Eq.~\eqref{eq:cos2thetaExp}a and exhibits substantially reduced uncertainty; it  encompasses both theoretical values, $b^{({\rm G})}$ and $b^{({\rm P5})}$, within its $1\,\sigma$ range.

We thus  adopt the conjecture that the Cabibbo angle arises as the smaller fraction in a golden or para-golden division of the solar mixing angle (though the  possibility of a distinct, numerically similar relation cannot be a priori excluded). This division rule mirrors the structure of Eq.~\eqref{eq:DefGolden}a and Eq.~\eqref{eq:GoldenProportionP5}a.

Accordingly, we obtain two predicted values for the Cabibbo angle, depending on which value of the small fraction parameter is used
\begin{equation}\label{eq:Ratio12GP5}
\theta_{\rm C}^{({\rm P5G})} = \theta_{12}^{l({\rm P5})} \, b^{({\rm G})} \approx 12.86^{\circ} \;\; ({\rm a}) \qquad
\theta_{\rm C}^{({\rm P5})}  = \theta_{12}^{l({\rm P5})} \, b^{({\rm P5})} \approx 12.97^{\circ} \;\;({\rm b}),
\end{equation}
where the angle given by  Eq.~\eqref{eq:Ratio12GP5}a is constructible.
The prediction~\eqref{eq:Ratio12GP5}b, is experimentally favoured  as the link between lepton and quark sectors.

As an independent cross-check of Eq.~\eqref{eq:GoldenRelations3}a, we evaluate the following expression using the measured value of the small fraction parameter, given by~\eqref{eq:Ratio12Exp}
\begin{equation}\label{eq:ThetaCSolarRatioSum}
b^{ql({\rm exp})} + \frac{1}{b^{ql({\rm exp})}}
= \frac{\theta_{\rm C}^{({\rm exp})}}{\theta_{12}^{l({\rm exp})}}
+ \frac{\theta_{12}^{l({\rm exp})}}{\theta_{\rm C}^{({\rm exp})}}
= 2.98 \pm 0.06,
\end{equation}
yielding a  tighter   agreement with the expected value of $3$, compared to~\eqref{eq:GoldenRelations3exp}a,  thus supporting the underlying division pattern.

Given the Cabibbo mixing angle, we can predict the second quark mixing angle by determining  a value for the mixing seed in the first quark pair, $\alphaC$. According to the results presented in Table~\ref{tab:Quantities}, the value of this  angle corresponds to the heptadecagon and is consistent with the mixing seed in the second lepton pair, justifying the previously adopted nomenclature as the Cabibbo mixing seed~\eqref{eq:DefCabibboSeed}. Thus, we obtain two, slightly different  predictions, depending on which value of the Cabibbo angle is used~\eqref{eq:Ratio12GP5}
\begin{equation}\label{eq:Theta23q}
\theta_{23}^{q({\rm P5G})} = \theta_{\rm C}^{({\rm P5G})} \tan \alphaC \approx 2.405^{\circ} \;\; ({\rm a})  \qquad
\theta_{23}^{q({\rm P5})}  = \theta_{\rm C}^{({\rm P5})} \tan \alphaC \approx 2.424^{\circ} \;\; ({\rm b}).
\end{equation}
Here, the angle $\theta_{23}^{q({\rm P5G})}$ is constructible, whereas the angle labelled 'P5' is not. Both predictions are in agreement with the measured value within $1\,\sigma$ uncertainty.

Since $\theta_{\rm C} = R_{\rm C} \cos \alpha_{\rm C}$, we can also predict the Cabibbo mixing strength parameter $R_{\rm C}$, yielding
\begin{equation}\label{eq:R1q}
R_{\rm C}^{({\rm P5G})} \approx 0.2284 \;\; ({\rm a}) \qquad
R_{\rm C}^{({\rm P5})}  \approx 0.2302 \;\; ({\rm b}).
\end{equation}
Comparison with the measured value, $R_{\rm C}^{({\rm exp})} = 0.23078 \pm 0.00070$ (Table~\ref{tab:Quantities}), again favours the para-golden 'P5' scheme.

The smallest quark mixing angle, $\theta_{13}^{q}$, can be similarly predicted by determining a value for the flavour mixing seed in the second quark pair, $\alpha_{2}^{q}$. The experimental value, $\alpha_{2}^{q({\rm exp})} = (5.10 \pm 0.16)^{\circ}$, does not allow for a unique identification of  the  corresponding polygon, despite its $3\%$ precision. However, we make use of the  result $\alpha_{2}^{q({\rm exp})}/\alpha_{\rm C}^{({\rm exp})}=0.488\pm 0.018$,  suggesting the ratio
$ \alpha_{2}^{q} / \alphaC = 1/2$, which aligns with the idea of constructibility as this choice  corresponds to the regular triacontatetragon ($34$-gon), which is constructible according to Wantzel’s theorem~\eqref{eq:Wantzel} with $k = 1$.

Under this assumption, $\theta_{13}^{q}$ can be expressed in two equivalent ways
\begin{equation}\label{eq:Theta13Quark}
\theta_{13}^{q} = \thetaC \tan \alphaC \tan \left( \frac{\alphaC}{2} \right) \;\; ({\rm a}) \qquad
\theta_{13}^{q} = R_{1}^{q} - \thetaC = \sqrt{\thetaC^2 + \left( \theta_{23}^{q} \right)^2} - \thetaC \;\; ({\rm b}).
\end{equation}
Eq.~\eqref{eq:Theta13Quark}b explicitly shows the constructibility of $\theta_{13}^{q}$ as the difference between the hypotenuse $A_1A_3$ and the cathetus $A_2A_3$ of the mixing triangle (Fig.~\ref{fig:PolygonGeneral}), provided the constructible $\thetaC$ is chosen~\eqref{eq:Ratio12GP5}a.

Using Eq.~\eqref{eq:Theta13Quark}a, the predicted values of the smallest quark mixing angle are
\begin{equation}\label{eq:Theta13Values}
\theta_{13}^{q({\rm P5G})} \approx 0.2228^{\circ} \;\; ({\rm a}) \qquad
\theta_{13}^{q({\rm P5})}  \approx 0.2247^{\circ} \;\; ({\rm b}),
\end{equation}
based on $\thetaC$ values from Eqs~\eqref{eq:Ratio12GP5}a and~\eqref{eq:Ratio12GP5}b, respectively. When the measured values are substituted  into Eq.~\eqref{eq:Theta13Quark}b, the result is $(0.219 \pm 0.008)^{\circ}$, within $1\,\sigma$ of  the independent measurement of $\theta_{13}^{q({\rm exp})}$.
The prediction for the mixing strength in the 2nd quark pair is  $\theta_{\rm C} \tan\alpha_{\rm C}/\cos\alpha_{\rm C} \approx 0.0424$, in a very good  agreement with the measurement.

Analogously to the lepton sector, the quark mixing angles can also be predicted as fractions of the straight angle, depending on  the choice of the small fraction parameter, $b$,~\eqref{eq:DefGolden}a or~\eqref{eq:GoldenProportionP5}a
\begin{equation}\label{eq:AnglesQuarksPi}
\theta_{\rm C} =_{\varepsilon} \pi \, b \tan \alphaC \;\; ({\rm a}) \quad
\theta_{23}^{q} =_{\varepsilon} \pi \, b \tan^2 \alphaC \;\; ({\rm b}) \quad
\theta_{13}^{q} =_{\varepsilon} \pi \, b \tan^2 \alphaC \tan \left( \frac{\alphaC}{2} \right) \;\; ({\rm c}).
\end{equation}

\section{Quark--Lepton Complementarity}\label{sec:QLC}

The derivation of the fundamental Quark--Lepton Complementarity (QLC) relation relies on the following nonobvious trigonometric $\varepsilon$-equality, which connects the regular pentagon and heptadecagon via their respective seed angles. It holds with exceptionally high accuracy, of order $10^{-5}$
\begin{equation}\label{eq:QLCcosPair1}
\cos \alpha_{\rm P5} + \cos \alpha_{\rm C} =_{\varepsilon} \frac{\pi}{2}.
\end{equation}

In terms of the mixing angles, according to Eqs~\eqref{eq:DefNormal1}--\eqref{eq:DefNormal2} and recalling that $\theta_{12({\rm n1})}^{l} = \theta_{12}^{l}$, this becomes
\begin{equation}\label{eq:QLCI}
\theta_{12}^{l({\rm P5})} + \theta_{\rm C(n)} =_{\varepsilon} \frac{\pi}{2}.
\end{equation}

In subsequent considerations we will need the experimental values of the  following normalised mixing angles
\begin{equation}\label{eq:TwoNormExp1}
\theta_{\rm C(n)}^{\rm exp}=(56.347\pm0.035)^{\circ}  \;\; ({\rm a}) \qquad \theta_{23({\rm n2})}^{l({\rm exp})}=(56.31\pm0.12)^{\circ}  \;\; ({\rm b}).
\end{equation}
Since $\alpha_{1}^{q} = \alpha_{2}^{l} = \alpha_{\rm C}$,  the following  equality   follows from Eqs~\eqref{eq:ParamPolar1}--\eqref{eq:ParamPolar2}
\begin{equation}\label{eq:QLCIIandIII}
\theta_{\rm C({\rm n})} = \theta_{23({\rm n2})}^{l},
\end{equation}
and  holds to within  a per mille when the experimental values~\eqref{eq:TwoNormExp1} are compared.

Substituting the experimental values into the left-hand side of~\eqref{eq:QLCI} yields $(90.04 \pm 0.73)^{\circ}$, in a very good agreement with a  right angle. This relation echoes certain QLC expressions inspired by GUT frameworks (see Sec.~\ref{sec:Introduction}). The geometric picture underlying~\eqref{eq:QLCI} involves normalising the heptadecagon to unit side length.

Moreover, Eq.~\eqref{eq:QLCI} implies the relation $\cos 2\theta_{\rm C(n)} =_{\varepsilon} - \cos 2\theta_{12}^{l({\rm P5})}$,
which enables a much more precise experimental  determination of the small fraction parameter. Using the value~\eqref{eq:TwoNormExp1}a, we get
\begin{equation}\label{eq:cos2thetaExpQ}
b_{\rm C}^{({\rm exp})} = -\cos 2\theta_{\rm C(n)}^{({\rm exp})} = 0.3858 \pm 0.0011,
\end{equation}
which is in excellent agreement with $b^{l({\rm exp})}$~\eqref{eq:cos2thetaExp}a. It also lies within less than $1\,\sigma$ of the para-golden value $b^{({\rm P5})}$~\eqref{eq:GoldenProportionP5}.
This highly accurate structure suggests that quark and lepton mixing angles may be connected,  provided a suitable normalisation is imposed.

We can also apply~\eqref{eq:QLCcosPair1} to  the lepton sector alone
\begin{equation}\label{eq:QLCanglesPair2L}
\theta_{12}^{l({\rm P5})} + \theta_{23({\rm n2})}^{l({\rm P5})} =_{\varepsilon} \frac{\pi}{2}
\end{equation}
and when the experimental values are substituted one gets $(90.00 \pm 0.74)^{\circ}$, again in close agreement with a right angle.

\section{Weak mixing and electroweak coupling constants}

In this section we address the issues pertaining to the  origin of the observed value of the Weinberg angle which, to date, remains largely unexplored, as well as  the concept of the corresponding complementarity, hereafter referred to as the Weinberg--Quark--Lepton Complementarity (WQLC).

A potential empirical connection between the Weinberg angle and the quark and lepton sectors can best be explored through the normalised mixing angles or, equivalently, the trigonometric functions of the mixing seed angles that  proved effective in  formulating the QLC relations, cf.~Eqs~\eqref{eq:QLCI} and~\eqref{eq:QLCIIandIII}. In this spirit, we observe that the following two independent experimental ratios yield a common value, consistent  within per mille accuracy~\eqref{eq:TwoNormExp1}, as expected from~\eqref{eq:QLCIIandIII}
\begin{equation}\label{eq:RatioThetaWThetaNorm}
\frac{\theta_{\rm W}}{\theta_{\rm C(n)}^{({\rm exp})}} =  0.5003\pm 0.0003  \quad ({\rm a}) \qquad \frac{\theta_{\rm W}}{\theta_{23({\rm n2})}^{l({\rm exp})}} =  0.5007\pm 0.0011  \quad ({\rm b}).
\end{equation}
This strongly suggests the relation
\begin{equation}\label{eq:ThetaWQLCI}
\theta_{\rm W} = f_{\rm W} \cos \alpha_{\rm C},
\end{equation}
where $f_{\rm W} = 1/2$  holds exactly or to $\varepsilon$-accuracy. With this value, the right-hand side of~\eqref{eq:ThetaWQLCI} evaluates to $28.1601^{\circ}$, which differs by one per mille from the on-shell value.
This empirical finding indicates that the Weinberg angle may belong to the same geometric framework associated with the heptadecagon as the Cabibbo angle, and may also indirectly relate  to the lepton mixing angles through this connection. Accordingly, the Weinberg angle corresponds to the line segment $A_2A_3$ in a regular heptadecagon with side length one-half (Fig.~\ref{fig:PolygonGeneral}), with $f_{\rm W}$ acting as the mixing strength. This geometric adherence gives principal support to~\eqref{eq:ThetaWQLCI}.

We note that a related angle, associated with the cathetus $A_1A_2$ of the same mixing triangle, appears in this context, namely $f_{\rm W} \cos \alpha_{\rm C} \approx 5.26^{\circ}$. Within the Standard Model, no obvious candidate phenomenon---whether related to mixing or of a different nature---appears to correspond to this angle, thereby suggesting the need for deeper insight.

Consequently, we can write the following WQLC relation
\begin{equation}\label{eq:ThetaWQLC2}
\theta_{12}^{l({\rm P5})} +  2 \theta_{\rm W} =_{\varepsilon} \frac{\pi}{2},
\end{equation}
where the left-hand side of~\eqref{eq:ThetaWQLC2} evaluates to $(90.06 \pm 0.73)^{\circ}$ when the experimental values are substituted, in excellent agreement with the right-angle.

From~\eqref{eq:ThetaWQLC2} one can derive a compact expression for  $\sin^2 \theta_{\rm W}$
\begin{equation}\label{eq:ThetaWQLCsin2}
\sin^2 \theta_{\rm W} =_{\varepsilon} \frac{1}{2} \left(1 - \sin \theta_{12}^{l({\rm P5})} \right).
\end{equation}
Substituting the measured solar angle into~\eqref{eq:ThetaWQLCsin2} yields $0.2227 \pm 0.0053$, in excellent agreement with the known value~(Table~\ref{tab:MixingExp}).

Based on~\eqref{eq:ThetaWQLC2}, we define the  'P5' Weinberg angle in order to  implement  strict equalities in the  considerations hereafter
\begin{equation}\label{eq:ThetaWP5}
\theta_{\rm W}^{(\rm P5)} := \frac{\pi}{4} - \vartheta,
\end{equation}
where
\begin{equation}\label{eq:thetaWexplicit}
\vartheta  = \frac{1}{2} \sqrt{1 - \left(\frac{\varphi}{2}\right)^2} \approx 16.8388^{\circ}.
\end{equation}
This yields $\theta_{\rm W}^{({\rm P5})} \approx 28.1612^{\circ}$, differing from the on-shell value by only one per mille.
The subsequent analysis may likewise be performed using an $\varepsilon$-equivalent definition of the angle $\vartheta$,
given by $\vartheta_{\rm C} = \frac{\pi}{2} \tan \alpha_{\rm C}\approx 16.8239^{\circ}$~\eqref{eq:Quasi3}.

A noteworthy implication of~\eqref{eq:ThetaWP5} is that the Weinberg rotation matrix $\mathcal{R}$ can be written as a linear combination of two opposite rotations
\begin{equation}\label{eq:Decomposition1}
\mathcal{R}\left(\theta_{\rm W}^{({\rm P5})}\right) = \cos \vartheta \, \mathcal{R}\left( \frac{\pi}{4} \right) + \sin \vartheta \, \mathcal{R}\left(-\frac{\pi}{4} \right),
\end{equation}
where the coefficients obey the unit-norm condition. This decomposition suggests a simple underlying structure for electroweak mixing—a superposition of two opposing rotations with weights determined by geometry.

Equation~(\ref{eq:ThetaWP5}), derived for $f_{\rm W} = 1/2$~(\ref{eq:ThetaWQLCI}), sets the stage for a series of key results.
In the Standard Model, the ratio of gauge couplings is related to the Weinberg angle by $\tan \theta_W = g^{\prime}/g$. Applying the tangent function to both sides of~(\ref{eq:ThetaWP5}) yields
\begin{equation}\label{eq:Ratioggprime1}
\tan \theta_W^{({\rm P5})} = \frac{1 - t}{1 + t} = \frac{g^{\prime}}{g},
\end{equation}
where the parameter $t$ is defined in terms of the golden ratio
\begin{equation}\label{eq:Functiont}
t = \tan \vartheta \approx  0.3027.
\end{equation}

The gauge coupling constants can be derived from the known Standard Model relations linking the masses of the vector bosons and the vacuum expectation value (VEV) $v = (\sqrt{2} G_{\rm F})^{-1/2} \approx 246.22~\mathrm{GeV}$
\begin{equation}\label{eq:MassesWiZ}
M_{\rm W} = \frac{1}{2} g v  \quad ({\rm a}) \qquad
M_{\rm Z} = \frac{1}{2} \sqrt{g^2 + g^{\prime 2}} \; v  \quad ({\rm b}),
\end{equation}
where $G_{\rm F}$ is the Fermi constant. By substituting  the values of $g$ and $g^{\prime}$ from~\eqref{eq:MassesWiZ} into~(\ref{eq:Ratioggprime1}) we get
\begin{equation}\label{eq:Ratioggprime2}
\frac{g^{\prime}}{1 - t} \approx 0.5018 \pm 0.0003 \quad (\mathrm{a}) \qquad \frac{g}{1 + t} \approx 0.5011 \pm 0.0001 \quad (\mathrm{b}),
\end{equation}
suggesting again to consider a  common value,  $f_{\rm g} = 1/2$, consistent  at a per mille level, which we adopt in subsequent considerations. We thus express the gauge coupling constants  in terms of the golden ratio as follows
\begin{equation}\label{eq:gandgprime1}
g^{\prime} = \frac{1}{2}(1 - t) \quad (\mathrm{a}) \qquad g = \frac{1}{2}(1 + t) \quad (\mathrm{b}).
\end{equation}

A notable  observation is that the above factor of  $1/2$ arises naturally from an additional assumption, consistent with the spirit of the present framework
\begin{equation}\label{eq:Sumggprime}
g^{\prime} + g = 1,
\end{equation}
i.e., that the gauge couplings divide unity according to the ratio given by the Weinberg angle~(\ref{eq:Ratioggprime1}). This division is analogous to the role of the golden ratio in the fractional scheme of lepton flavour mixing.
As a consequence of Eq.~(\ref{eq:Sumggprime}), the vector boson masses can be written purely in terms of the Fermi constant (through $v$) and the golden ratio
\begin{align}
M_{\rm W} & = \frac{1}{4}(1 + t) \; v & \approx 80.1877~\mathrm{GeV} \quad (\mathrm{a}) \nonumber \\
M_{\rm Z} & = \frac{1}{2\sqrt{2}} \sqrt{1 + t^2}\; v \quad &\approx 90.9527~\mathrm{GeV} \quad (\mathrm{b}), \label{eq:Massesbyphi}
\end{align}
yielding  predictions that match  the measured values  at  a per mille level~(Table~\ref{tab:MixingExp}).

In the Standard Model,  the  electron charge, $e$,  is expressed  in terms of  the  gauge coupling constants;  by substituting~(\ref{eq:gandgprime1}),
we express it solely by $t$ as follows
\begin{equation}\label{eq:ElectronCharge1}
e = \frac{g g^{\prime}}{\sqrt{g^2 + g^{\prime 2}}} = \frac{1 - t^2}{2\sqrt{2} \sqrt{1 + t^2}} \approx 0.3074,
\end{equation}
from which the expression for the  fine-structure constant can be  derived
\begin{equation}\label{eq:FineStructureConstant1}
\alpha_{\rm em} = \frac{e^2}{4\pi} = \frac{(1 - t^2)^2}{32\pi (1 + t^2)} \approx 0.007533 \approx \frac{1}{133}.
\end{equation}
Thus the fundamental, dimensionless constant $\alpha_{\rm em}$ can  be expressed purely in terms of the golden ratio $\varphi$ or $\varepsilon$-equivalently in terms of the Cabibbo seed angle $\alpha_{\rm C}$ via $t$. Thus in  the present framework, it is   rooted in the geometry of  constructible polygons. The leading-order term  in Eq.~(\ref{eq:FineStructureConstant1}) evaluates to $1/(32\pi) \approx 1/100$, with the final value obtained through  higher-order corrections.

The above  considerations  suggest the emergence of two semi-empirical factors, $f_{\rm W}$ and $f_{\rm g}$, each equal or close to $1/2$. Allowing for deviations, we analyse Eq.~(\ref{eq:Sumggprime}) in full generality
\begin{equation}\label{eq:SumggprimeVariable}
f_{\rm g} \left[ \tan\left(f_{\rm W} \frac{\pi}{2}\right) - \tan\left(f_{\rm W} \sqrt{1 - \left(\frac{\varphi}{2}\right)^2} \right) + 1 + \tan\left(f_{\rm W} \frac{\pi}{2}\right) \tan\left(f_{\rm W} \sqrt{1 - \left(\frac{\varphi}{2}\right)^2} \right) \right] = 1.
\end{equation}
It is straightforward to verify that Eq.~(\ref{eq:SumggprimeVariable}) admits a unique rational solution
\begin{equation}\label{eq:AnthropicPoint}
f_{\rm W} = f_{\rm g} = \frac{1}{2},
\end{equation}
hereafter referred to as the \emph{anthropic point} on the $f_{\rm W}$--$f_{\rm g}$ plane.
This solution is special in that $f_{\rm W} = 1/2$ defines a Weinberg angle constructible with compass and straightedge, and given  Eq.~(\ref{eq:Sumggprime}), this necessitates $f_{\rm g} = 1/2$. Any deviation from the anthropic point along the curve~(\ref{eq:SumggprimeVariable}) alters the gauge and electromagnetic couplings and may lead to values prohibiting the emergence of the Universe and  life as we know it. Assuming a maximal acceptable variation of $\theta_W$  amounting to  $\sim 1^{\circ}$  implies that $f_{\rm W}$ and $f_{\rm g}$   remain roughly  within   $(2$--$3)\%$ of the anthropic point, given the approximate linear relation $f_{\rm g} \approx 1 - f_{\rm W}$.

\section{Summary and Outlook}\label{sec:Conclusions}

We  have  presented a novel, semi-empirical  framework for describing flavour mixing at the low energy scale, based on Euclidean geometry and number theory---specifically, simple constructs associated with regular polygons that are compass-and-straightedge constructible, namely the pentagon and the heptadecagon, which are linked to the Fermat primes.  The golden ratio emerges naturally as a geometric consequence once the pentagon is involved. A parallel fractional scheme in the lepton sector, based on the unit line segment division rule, suggests  that  the golden ratio consistently  governs the normalisation across sectors through the solar and the Cabibbo angles. The  values of the  quark  and lepton mixing angles are predicted with unprecedented  accuracy.

Intriguingly, the Weinberg angle aligns with this same geometric framework, suggesting that both electroweak and flavour mixing may be described by  the same elementary geometric structure at the present energy scale.  Thus the Bi-Constructible pattern offers a semi-empirical unification of weak and flavour mixing without resorting to  abstract symmetries.  The proportionality of the  electroweak gauge couplings to VEV, elementary  charge  and  the fine-structure constant can be expressed in terms of the golden ratio (or $\varepsilon$-equivalently, the Cabibbo seed angle) with remarkable accuracy, raising the possibility that, in  particular,  $\alpha_{\rm em}$  may emerge   from  pure arithmetic or  geometric constraints in Nature.

Our framework points toward a hidden simplicity behind mixing phenomena at the present energy scale---one that is geometric, number-theoretic, and potentially anthropic in nature. Whether, and possibly how,  these low-energy structures are connected to high-energy-scale symmetries remains an open question.
Such a scenario is, a priori, natural to expect though it carries a non-trivial caveat.
On the one hand, the appearance of the pentagon and the golden ratio may suggest a   connection to a specific flavour symmetry,  like those mentioned  in Sec.~\ref{sec:Introduction}.  On the other hand, the heptadecagon does not appear to be  linked  to any known flavour symmetry or discrete group typically considered in particle physics, aside from the  dihedral group $D_{17}$, unconventional in flavour model building. This observation rises a question whether   the Bi-Constructible scheme is  intrinsic to the underlying geometric structure itself, rather than imposed through group-theoretical assumptions---an issue  deserving  further exploration.
A related question concerns the role of constructibility in the appearance of specific polygons in weak and flavour mixing---whether this feature is a mere coincidence or a purposeful aspect of Nature. In passing, the alignment between the particle content of the Standard Model, 5 bosons and 17 (bosons and fermions) in total, and the Fermat primes, though possibly coincidental, is also puzzling.

Lastly, the empirical part of  this study was partly limited by the uncertainty in the atmospheric mixing angle, for which a conservative value was adopted due to tensions among current best-fit results. A reliable average from direct measurements remains essential to further refine  the framework. The quark sector, while more stable, still requires improved precision in CKM matrix elements, as small inconsistencies among determinations  result  in increased  uncertainties due to rescaling.

\bibliography{MixingAnglesFinal1}

\end{document}